\documentclass[english,journal=jctcce,manuscript=article,email=true,etalmode=truncate,maxauthors=0,doi=true]{achemso}

\usepackage{graphicx}
\graphicspath{{figures/}}
\usepackage{physics}
\usepackage{multirow}
\usepackage{xcolor}
\usepackage{caption}
\usepackage{subcaption}
\usepackage{xcolor}
\usepackage{bm}
\usepackage{eufrak} 
\usepackage{yfonts} 
\usepackage[title,titletoc]{appendix}
\usepackage{diagbox}
\usepackage{multicol}

\usepackage[numbers,sort&compress,super]{natbib}
\usepackage{natmove}
\setkeys{acs}{doi = true}

\usepackage{hyperref} 
\hypersetup{colorlinks=true, citecolor=blue, urlcolor=blue, linkcolor=blue}
\usepackage{cleveref}
  	\crefname{figure}{Figure}{Figures}
  	\crefname{table}{Table}{Tables}
  	\crefname{equation}{Eq.}{Eqs.}
  	\crefname{section}{Section}{Sections}
  	\crefname{subsection}{Section}{Sections}
  	\crefname{subsubsection}{Section}{Sections}
  	\crefname{algorithm}{Algorithm}{Algorithms}

\newcommand{\doi}[1]{\href{https://dx.doi.org/#1}{\nolinkurl{#1}}}

\usepackage[section,frozencache]{minted}	

\newcommand{\code}[1]{\texttt{#1}}
\newcommand{\ga}{\ensuremath{\mathbf{a}}}
\newcommand{\gb}{\ensuremath{\mathbf{b}}}
\newcommand{\gc}{\ensuremath{\mathbf{c}}}

\newcommand{\gC}{\ensuremath{\textswab{c}}}

\newcommand{\gn}{\ensuremath{\mathbf{n}}}

\newcommand{\gtt}{\ensuremath{\mathbf{t}}}
\newcommand{\gu}{\ensuremath{\mathbf{u}}}
\newcommand{\gzero}{\ensuremath{\bm{0}}}

\usepackage{todonotes}

\title{Memory-Efficient Recursive Evaluation of 3-Center Gaussian Integrals} 

\author{Andrey Asadchev}

\author{Edward F. Valeev}
\affiliation{Department of Chemistry, Virginia Tech, Blacksburg, VA 24061}
\email{efv@vt.edu}

\begin{document}

\date{\today}

\begin{abstract}
To improve the efficiency of Gaussian integral evaluation on modern accelerated architectures  FLOP-efficient Obara-Saika-based recursive evaluation schemes are optimized for the memory footprint. For the 3-center 2-particle integrals that are key for the evaluation of Coulomb and other 2-particle interactions in the density-fitting approximation the use of multi-quantal recurrences (in which multiple quanta are created or transferred at once) is shown to produce significant memory savings. Other innovation include leveraging register memory for reduced memory footprint and direct compile-time generation of optimized kernels (instead of custom code generation) with compile-time features of modern C++/CUDA. Performance of conventional and CUDA-based implementations of the proposed schemes is illustrated for both the individual batches of integrals involving up to Gaussians with low and high angular momenta (up to $L=6$) and contraction degrees, as well as for the density-fitting-based evaluation of the Coulomb potential. The computer implementation is available in the open-source \code{LibintX} library.
\end{abstract}

\maketitle 

\section{Introduction}\label{sec:intro}

Evaluation of Gaussian integrals\cite{VRG:boys:1950:PRSMPES,VRG:gill:1994:AiQC,VRG:reine:2011:WIRCMS} accounts for a significant or a dominant portion of the total cost of many key tasks in Gaussian LCAO electronic structure computations of molecules and solids. Therefore efficient evaluation of various operators in Gaussian AO bases --- and in particular, 2-body Coulomb integrals (i.e., the electron repulsion integrals) --- has been the focus of much attention of the electronic structure community,\cite{VRG:boys:1950:PRSMPES,VRG:shavitt:1963:QM,VRG:dupuis:1976:JCP,VRG:mcmurchie:1978:JCP,VRG:pople:1978:JCP,VRG:rys:1983:JCC,VRG:obara:1986:JCP,VRG:head-gordon:1988:JCP,VRG:augspurger:1990:JCC,VRG:gill:1990:JPC,VRG:gill:1991:IJQC,VRG:ishida:1991:JCP,VRG:hamilton:1991:CP,VRG:lindh:1991:JCP,VRG:ishida:1993:JCP,VRG:johnson:1993:CPL,VRG:ten-no:1993:CPL,VRG:ishida:1996:IJQC,VRG:bracken:1997:IJQC,VRG:adams:1997:JCP,VRG:valeev:2000:JCP,VRG:dupuis:2001:JCP,VRG:weber:2004:CPC,VRG:ahlrichs:2004:PCCPP,VRG:kobayashi:2004:JCP,VRG:ahlrichs:2006:PCCPP,VRG:makowski:2007:IJQC,VRG:flocke:2008:JCC,VRG:ishimura:2008:TCA,VRG:shiozaki:2009:CPL} with important developments continuing unabated.\cite{VRG:asadchev:2012:CPC,VRG:sandberg:2012:JCP,VRG:rak:2015:CPL,VRG:pritchard:2016:JCC,VRG:samu:2017:JCP,VRG:kalinowski:2017:JCTC,VRG:golze:2017:JCP,VRG:kussmann:2017:JCTC,VRG:hayami:2018:IJQC,VRG:tornai:2019:JCTC,VRG:peels:2020:JCTC,VRG:neese:2022:JCC}

A particular challenge for the electronic structure community has been the greatly expanded importance of the data parallelism for the performance of modern processors. Compared to the other key kernels of the electronic structure, namely, the linear and tensor algebra, evaluation of Gaussian integrals is difficult to optimize due to many factors; among the most important are: (1) the relatively low arithmetic intensity of the Gaussian integral kernels, (2) their irregular computation and data access patterns, and (3) significant dependence of the distributions of shell-set costs and sizes on the AO basis set family and cardinal rank (such as $X$ in the correlation-consistent basis set family cc-pV$X$Z). All of these factors make it especially challenging to port Gaussian integral kernels onto accelerated co-processors, such as general-purpose graphical processing units (GPGPUs, or, simply, GPUs), that have become the norm both on the commodity and high-end platforms. Hence there has been an intense effort to address these challenges, both on the modern central processing units (CPUs) with wide single-instruction-multiple-data (SIMD) instructions\cite{VRG:pritchard:2016:JCC} and on GPUs.\cite{VRG:yasuda:2007:JCC,VRG:ufimtsev:2009:JCTC,VRG:yasuda:2014:IJQC,VRG:rak:2015:CPL,VRG:kalinowski:2017:JCTC,VRG:song:2016:JCTC,VRG:kussmann:2017:JCTC,VRG:tornai:2019:JCTC,VRG:barca:2020:JCTC,VRG:barca:2020:2SICHPCNSAS,VRG:barca:2021:JCTC,VRG:barca:2021:PICHPCNSA,VRG:johnson:2022:JCTC}

In this work we design an efficient approach for evaluation of 3-center 2-body Gaussian integrals on massively-data-parallel devices like modern GPUs. The decision to focus on 
3-center 2-body integrals is due to their
foundational role in the density fitting  technology\cite{VRG:whitten:1973:JCP,VRG:baerends:1973:CP,VRG:dunlap:2000:PCCPP} that is crucial for efficient evaluation of many-body operators in electronic structure.\cite{VRG:weigend:2002:JCP,VRG:manby:2003:JCP,VRG:neese:2009:JCP,VRG:kelley:2013:JCP,VRG:merlot:2013:JCC,VRG:hollman:2014:JCP,VRG:manzer:2015:JCP,VRG:tew:2018:JCP,VRG:wang:2020:JCP} The density fitting technology is especially crucial for the electronic structure on GPUs by trading floating-point operations (FLOPs) for reduced memory footprint; this makes DF a perfect companion for the modern memory-limited {\em FLOP-rich} compute devices. While our work is specific to 3-center evaluation strategies\cite{VRG:ahlrichs:2006:PCCPP}, the main ideas apply directly to 4-center Gaussian integrals.
Lastly, while some implementation details of our work are specific to the particular programming model of GPUs considered here (CUDA), the key algorithmic innovations can be exploited on other data-parallel devices like modern SIMD-capable CPUs.

The rest of the manuscript is organized as follows. \cref{sec:analysis} discusses the 3-center integral evaluation in the context of modern GPU architectures and their programming models; the conclusion is that efficient recursive evaluation of 3-center Gaussian integrals is possible on modern GPUs by reducing the memory footprint to fit entirely inside the ``fast'' (shared) memory. \cref{sec:implementation} discusses crucial implementation details, such as how the Gaussian integral recurrences can be implemented entirely in modern C++, without the need for a specialized code generator, as well as brief details about the user API of \code{LibintX}. \cref{sec:performance} reports the performance of our integral engine on conventional CPUs and NVIDIA's V100 devices for evaluation of individual integrals as well as for evaluation of the Coulomb potential matrix. \cref{sec:summary} summarizes out findings and outlines the next steps. The notation used throughout this paper is defined in Appendix \ref{sec:notation}.

\section{\label{sec:analysis} Analysis}

Our objective is to design a single evaluation strategy capable of competitive (even if not optimal) performance for integral classes with $L$ up to 6 and varying contraction degrees and optimized for modern and future heterogeneous platforms. To motivate the choice of a particular evaluation method we first must review the basics of the relevant aspects of the GPU architecture and programming models (\cref{sec:gpuoverview}); due to the space limitations the reader is referred to the respective hardware and programming model manuals for more details.  Evaluation, design and implementation strategies are then discussed in \cref{sec:rysornot,sec:recurrence,sec:codegen}.

\subsection{Overview of GPU programming models and architecture}\label{sec:gpuoverview}

Although there exist several models for programming GPUs and other accelerators, our work focuses on NVIDIA's CUDA programming model, as it is the most established programming model based on the C++ programming language (the importance of C++ for our purposes will become clear later). Other vendors' programming models for data-parallel processors (HIP, DPC++), as well as the multi-vendor SYCL programming model, are modeled closely after CUDA.
Thus porting CUDA code to other accelerator architectures should be relatively straightforward.

A single-process CUDA program consists of one or more threads of execution on the host inserting device code (CUDA {\em kernel}) invocations into one or more CUDA {\em streams}. Each stream executes kernels sequentially (in-order) but kernels from multiple streams can execute at the same time; thus CUDA streams are analogous to the threads of a thread pool on the host. Inserting and scheduling a CUDA kernel invocation involves substantial overhead even when single host thread is involved, on the order of a few microseconds; in such a short period of time a modern device can execute up to a 100 MFLOPs, thus the amount of work per device kernel invocation must substantially exceed 1 GFLOPs. For the sake of managing  the code complexity the device kernels can include calls to other device functions (to avoid confusion with CUDA kernels that are ``invoked'' from the host code, we will refer to them here as {\em subkernels}) which are  often inlined by the compiler, hence the effective cost of subkernel calls is negligible.

Key concepts of the CUDA execution model are {\em threads} and {\em thread blocks}. Execution of an instruction by a thread is analogous to executing a single scalar component of a vector (SIMD) instruction by a CPU. Threads are scheduled in {\em blocks}, with each thread block further internally partitioned into a set of atomic groups of threads ({\em warps}); the warps of a single thread block are typically executed concurrently. Each thread block is bound to a {\em streaming multiprocessor} (SM), which is analogous to 
a single CPU core.  Each SM may be executing warps from one or more thread blocks concurrently.  Having multiple thread blocks resident on an SM allows to hide latency of certain operations, such as the main memory accesses.

The CUDA memory hierarchy\cite{VRG:jia:2018:AC} includes  {\em registers} (private to a thread), {\em shared memory} (private to a thread block), and {\em global memory} (accessible from any thread). These memory spaces correspond to hardware memories located on each SM (registers, shared memory, L1 cache of the global memory) and shared by all SMs (L2 and optionally higher level caches and DRAM).

A distinctive feature of modern GPUs is the availability of per-SM shared memory, also known as {\em local data store} (LDS) in ROCm/HIP and {\em local memory} in SYCL; {\em scratchpad} memory is a general term that is often used to describe these types of memory. Several properties of shared memory make it the optimal location for non-register data in a high-performance code on GPU: (a) its low latency (up to $50\times$ lower than that of a location in main memory missing from the Translation Lookaside Buffer (TLB)\cite{VRG:jia:2018:AC}), (b) usually fast nonsequential access from consecutive threads (nonsequential accesses to the main memory can be hindered by coalescing), and (c) fast reads/writes relative to the main memory. Although the shared memory must be managed explicitly, that is an advantage for the high-performance code.

These favorable features of the shared memory thus motivate the central objective of the current work: to design an integral evaluation strategy that ensures that the entire data can fit into the registers and/or shared memory. Although the total size of registers and shared memory varies between devices and architectures, the typical amount is on the order of a few 100s of kB of memory. For example, each SM on the NVIDIA V100 GPU has 256 KiB of registers and up to 96 KiB of shared memory, while the newer NVIDIA A100 GPU has up to 160 KiB of shared memory per SM. These figures are in line with the corresponding hardware characteristics of high-end GPUs from other vendors.
Also note that these numbers are per SM, not per thread block: to allow SM concurrency each thread block must use at most half of the shared memory and registers.

Making sure that only the shared memory and registers are utilized is a tall task for a Gaussian integral code, even if we restrict ourselves to 3-center integrals only. For example, the $(ii|\textswab{i})$  shell-set (i.e., Cartesian bra Gaussians and a solid-harmonic ket Gaussian) alone occupies ~80 kB, thus trying to compute it purely within the registers and shared memory requires several innovations that we describe below.

\subsection{Design Pivot 1: Quadrature or Recurrence?}\label{sec:rysornot}

Our design of an efficient evaluation method of a 3-center 2-body integral over contracted Gaussians started with the following assumptions:
\begin{itemize}
\item the ket Gaussian will always be a solid harmonic Gaussian rather than a Cartesian Gaussian; i.e. our targets are $(\ga \gb|\gC)$ integrals;
\item each device kernel will evaluate multiple target shell-sets of a given type to ensure enough work to offset the overhead of the kernel launch from the host;
\item each shell-set will be evaluated in a single thread block, with parallelism over integrals and/or primitives (contrast such {\em intra}-shell-set parallelism to parallelization over multiple shell-sets in Ref. \citenum{VRG:pritchard:2016:JCC});
\item integrals over Gaussians with $L$ up to 6 with low and high $K$ are targeted.
\end{itemize}
Our primary target is evaluation of integrals in using double precision  floating point arithmetic (FP64) on high-end GPU devices of today, namely the NVIDIA V100 card.

We first considered the Rys quadrature.\cite{VRG:dupuis:1976:JCP,VRG:rys:1983:JCC} General consensus is that it is the most memory efficient algorithm for high angular
momentum integrals. However, its FLOP counts are significantly suboptimal for low-$L$ high-$K$ integrals.\footnote{Note that the total FLOP count is not necessarily a good performance model for complex codes like integral kernels.} Rys quadrature is also more difficult to generalize to
non-Coulomb integrals compared to the Boys function-based recursive schemes.\cite{VRG:ahlrichs:2006:PCCPP} Most importantly, it is not as well suited for evaluation of high angular momentum integrals as we believed. Consider $(hh|\textswab{i})$ shell-set. In addition to $21^2 \times 13=5733$ words
($\approx 46$ kB in FP64) needed to hold the
result, the direct (HRR-free) Rys strategy requires
storing 9 (number of roots) $\times$ 3 (number of Cartesian axes)
$ \times 6^2 \times7 $ 1-d integrals, or $\approx 54$ kB.
Without resorting to batched evaluation this requires 100 kB of shared memory to launch just 1 thread block per SM; this exceeds the shared memory capability of all but the latest high-end NVIDIA A100 device.

\subsection{Design Pivot 2: Recurrences (Which and How)}\label{sec:recurrence}

We next considered the McMurchie-Davidson method,\cite{VRG:mcmurchie:1978:JCP} a popular choice for CPU\cite{VRG:neese:2022:JCC} and GPU
implementations\cite{VRG:ufimtsev:2009:JCTC} due its relative simplicity compared to other recursive schemes. Unfortunately the FLOP counts of the MD method
are significantly higher than those of the Obara-Saika-based approaches.
Although we did not pursue a full McMurchie-Davidson
CUDA implementation for the integral evaluation, we leveraged it for the Coulomb potential evaluation (J-Engine) that will be discussed elsewhere.

Lastly we turned our attention to the Obara-Saika-based recurrences.\cite{VRG:obara:1986:JCP,VRG:head-gordon:1988:JCP,VRG:ahlrichs:2006:PCCPP}
The Obara-Saika-based methods seemed to us to be a poor choice for GPUs:
the recurrences are ``wide'' (i.e., typically involve more than 2 terms on the right-hand side), are nonuniform, and require intermediate storage that (unlike the Rys quadrature) significantly exceeds the size of the target shell-set itself. Furthermore, implementation of Obara-Saika schemes is complicated and usually calls for an optimizing compiler like our \code{Libint}.\cite{VRG:valeev:2021:libint-2.7.0} Due to the assumption of a solid-harmonic ket Gaussian, it is possible to use the Ahlrichs' simplification\cite{VRG:ahlrichs:2004:PCCPP} of the Head-Gordon-Pople (HGP) scheme,\cite{VRG:head-gordon:1988:JCP} however, that alone is not sufficient to address the main concerns.

Several key ideas allowed us to evaluate the integrals efficiently within the OS framework:
\begin{itemize}
\item {\bf In-place evaluation}. CPU-only kernels implementing integral recurrences (such as in \code{Libint}) usually consists of sequences of function (subkernel) calls each of which evaluates a shell-set from two or more input shell-sets\footnote{\code{Libint} can generate code optimized at the level of individual integrals, rather than shell-sets, with the kernel expressed as a single function; this approach is used only for low-angular momentum integrals, but is essential for optimization of especially geometric derivatives of integrals}.
Due to the complexity of the integral recurrences it is not in general possible to write the resulting integrals into the memory occupied by the inputs. Thus the output of such a subkernel must be written to a  memory segment that is disjoint from those of the inputs. Modern GPUs's allow to reduce the peak memory consumption of such subkernels by leveraging their massive register file (256 KiB per SM on the NVIDIA V100 card compared to merely few dozen \{1,2\} KiBs per each \{AVX2,AVX512\} x86\_64 CPU core) that significantly exceeds the size of the shared memory. Due to the ability to {\em control programmatically} the data transfers between the registers and the shared memory it is possible to evaluate the complete output shell-set in registers and then overwrite the shared memory occupied by the recurrence inputs by the corresponding output at the end of the subkernel. Specifically, the input integrals can be fully read into the registers and the extra registers will be still available to compute one {\em or more} rounds of recursion. Then the results of the recursion can be committed to the shared memory in the order needed in the next stage. The only limitation of this technique is that each computation stage between the consecutive shared memory reads/writes can be mapped onto the register file at compile time. 
\item {\bf {\em Multiquantal} recurrences}.  Integral recurrences are typically expressed in a form where single quantum is built up or transferred at a time. Multiple application of a given recurrence can be viewed as a multiquantal form of the recurrence in which multiple quanta are built up or transferred at a time. 
Their use has been shown to decrease the number of memory operations, that was postulated by Frisch et al.\cite{VRG:frisch:1993:CPL} to be a better important objective function for optimizing the Gaussian integral kernels than the traditional total FLOP count, in the context of HRR.\cite{VRG:ryu:1991:CPL,VRG:johnson:1993:CPL,VRG:makowski:2007:IJQC} Here we show that not only memory operations are reduced by the use of multiquantal relations, but the (shared) memory footprint is reduced also. For this reason it is beneficial to apply them not just to HRR, but to the VRRs also.  Although the use of multiquantal recurrences increases the operation (FLOP) count, this can be a worthy tradeoff on highly-parallel FLOP-rich memory-constrained GPU devices; others have postulated such notions in the context of GPU evaluation of algorithms as well.\cite{VRG:yasuda:2014:IJQC}.
\end{itemize}
It should be noted that these ideas are not limited to the Obara-Saika-based schemes, are are more broadly applicable.
Furthermore, our implementation is not yet fully optimal; future improvements will be reported in due time.

\subsection{Design Pivot 3: Code Generation (Custom or Generic)}\label{sec:codegen}

Most Gaussian integral engines use custom code generators to emit optimized kernels for computing integrals of specific class(es), in whole or in part.
This is especially so when it is necessary to manage complex recurrences like those in the Obara-Saika framework; due to the irregular structure of the Gaussian integral recurrences
their efficient implementation calls for code generation to eliminate
the dispatch, addressing, and resource management logic.
Another benefit of code generation is its ability to cover
the feature space of an integral engine efficiently, such as support
for integrals and their derivatives over dozens of quantum operators
utilized in electronic structure, integrals over Gaussian spinors, and
so on.\cite{VRG:sun:2015:JCC}

The level of sophistication of integral code generation\cite{VRG:womack:2015:phdthesis,VRG:song:2016:JCTC,VRG:kussmann:2017:JCTC,VRG:tornai:2019:JCTC} can range from ad hoc tools to embedded or standalone DSLs equipped with custom compilers (e.g., for \code{Libint} \cite{VRG:valeev:2021:libint-2.7.0}). This is an approach that has proven its worth over the decades of use in the real world, but is not free from drawbacks. 
First, code generation  decouples
processing of the integral evaluation logic from its implementation; i.e.,
programming a code generator is {\em metaprogramming}.
This makes development of a production tool for code generation a formidable challenge. Second, code generator produces kernel code in a high-level language that needs to be compiled further. The separate generation/compilation stages complicate any task crossing of the boundary between the code generator and generated code, e.g., when exploring performance, debugging, and overall maintenance. 

Here instead we pursue a simpler strategy: instead of a custom compiler we use the powerful {\em compile-time} capabilities of modern (2017 and later) C++ and its CUDA extension, namely template-based metaprogramming and compile-time expressions (\code{constexpr}), to generate integral kernels {\em and} compile them using the C++ compiler itself.
Some aspects of this approach have precedents; for example, \code{Libint} already generates some kernels using manually-written templates rather than explicit code generation by a custom compiler to reduce the compilation time of shell-set recurrence relation kernels for high $L$.  Here, however, the use of compile-time code generation is taken to new heights and used throughout to generate all kernels. While such an approach does not automate all optimization tasks, the overall simplification (lack of a custom compiler, improved maintainability and portability) make it at a minimum an interesting experiment in its own right and potentially sufficient for all practical purposes.

We emphasize that the compile-time approach to kernel generation is not targeted as a {\em replacement} for  the traditional custom compiler-based approach. Since compile-time programming in C++ (both template- and \code{constexpr}-based) is Turing-complete, it is therefore possible to use it to implement an arbitrary program, including a custom compiler for integrals, within the host language (C++) itself. Unlike the traditional code-generation approach, in the compile-time-only approach both optimization, kernel generation, and and compilation of the kernels to the native (machine) code happens within a single program. However, the compile-time-only approach would be awkward due to the facts that compile-time programs lack mutable state, hence implementation would be necessarily be less efficient than the runtime counterpart (i.e., a custom compiler). 
Hence here we use compile-time approach to primarily generate kernels for a manually-designed evaluation strategy, i.e., the optimization burden is largely borne by the programmer. However, even the custom compiler is limited in how much optimization it can automate due to (a) NP-completeness of optimization tasks (even for low-$L$ integral sets exhaustive optimization is not possible) and (b) limited usability of simple performance models (total FLOP count is not a good performance model on modern machines). Thus the runtime-time (custom compiler) and compile-time techniques (and their blends, as already demonstrated within \code{Libint}) are complementary implementation techniques for a generic integral compiler toolchain, both with strengths an weaknesses. No matter the implementation strategy design of high-quality heuristics and performance models (e.g., profile-guided optimization) must be addressed separately.

\subsection{Recursive Evaluation Formalism and Implementation \label{sec:impl}}

\subsubsection{Overview of the Obara-Saika-based approach \label{sec:hgpa}}
We use the Obara-Saika-based Head-Gordon-Pople-Ahlrichs (HGPA) approach.
The HGPA method\cite{VRG:ahlrichs:2004:PCCPP} for evaluation of $(\ga \gb|\gC)$ integrals over contracted Gaussians (with $\gC$ a solid harmonic Gaussian) involves 3 key identities:
\begin{align}
\label{eq:hrr}
    \text{HRR}:& (\ga \, \gb+\bm{1}_i|\gC) = & (\ga+\bm{1}_i \, \gb|\gC) + (\mathbf{AB})_i (\ga \, \gb|\gC), \\
\label{eq:vrr2}
    \text{VRR2}:& [\ga \, \gzero |\gc + \bm{1}_i]^{(m)} = &  \frac{\rho}{\zeta_c}(\mathbf{PC})_i[\ga \, \gzero |\gc]^{(m+1)} + \frac{a_i \rho}{2\zeta_c\gamma}[\ga - \bm{1}_i \, \gzero |\gc]^{(m+1)}, \\
\label{eq:vrr1}
    \text{VRR1}:& [\ga + \bm{1}_i \, \gzero |\gzero]^{(m)} = & (\mathbf{PA})_i [\ga \, \gzero |\gzero]^{(m)} -
    \frac{\rho}{\gamma}(\mathbf{PC})_i [\ga \, \gzero |\gzero]^{(m+1)} \nonumber \\
    & & + \frac{a_i}{2\gamma} [\ga - \bm{1}_i \, \gzero |\gzero]^{(m)} - \frac{a_i \rho}{2 \gamma^2} [\ga - \bm{1}_i \, \gzero |\gzero]^{(m+1)}.
\end{align}
Evaluation of the target shell-set $(a \, b|\gC)$ in HGPA proceeds as follows:
\begin{itemize}
    \item Integrals $[\gzero \, \gzero |\gzero]^{(m)}$, $c \leq m \leq \dots a+b+c$ are evaluated;
    \item Shell-sets $[e \, 0 |0]^{(c)}$,  $|a-c| \leq e \leq a+b$  are evaluated via VRR1;
    \item Shell-sets $[e \, 0 |c]$, $a \leq e \leq a+b$  are evaluated via VRR2;
    \item Shell-sets $[e \, 0 |c]$ are contracted to produce $(e \, 0 |c)$;
    \item The ket Gaussian in $(e \, 0 |c)$ is transformed from the Cartesian to the solid harmonic form to produce $(e \, 0 |\gC)$;
    \item Target shell-set $(a \, b|\gC)$ is evaluated from $(e \, 0 |\gC)$ via HRR.
\end{itemize}

\subsubsection{Evaluation of the \ensuremath{[\mathbf{0 \, 0} | \mathbf{0}]^{(m)}} integrals \label{sec:boys}}

For the Coulomb operator the $[\gzero \, \gzero |\gzero]^{(m)}$ integrals are obtained from the Boys function as
\begin{align}
    [\gzero \, \gzero |\gzero]^{(m)} = \frac{2 \pi^{5/2}}{\left(\gamma + \zeta_c\right)^{1/2} \gamma \zeta_c} \exp(-\frac{\zeta_a \zeta_b}{\zeta_a + \zeta_b} |{\bf AB}|^2) F_m(\rho |{\bf PC}|^2).
\end{align}
The Boys function values are
evaluated using the 7th-order Chebyshev interpolation (essentially, a
high-order extension of
Ref. \citenum{VRG:gill:1991:IJQCa}). The number and width of interpolation intervals and the maximum value of $m$ is constructed to guarantee better than FP64 epsilon
(in absolute {\em and} relative sense) everywhere in the interpolation
range. Computations are simple uniform polynomial evaluation and have
cache-friendly access patterns. Value of $F_m(x)$ for each primitive (i.e., value of $x$) and value of $m$ is computed by a single thread to maximize parallelism. For values of $x$ outside of the interpolation table range (namely, $x>117$) upward recursion in $m$ is used. The cost of the Boys function evaluation is insignificant, except for low-$L$ high-$K$ shell-sets.

\subsubsection{Implementation of VRR1\label{sec:impl-vrr1}}
Even for the largest, $(i \, i|\textswab{i})$, target shell-set all shell-sets produced by VRR1, namely $[e \, 0|0]^{(6)}$ with $0 \leq e \leq 12$, occupy 455 FP64 words, or approximately  $3.6$ KiB. Thus it may seem that it is not important to optimize VRR1 for the memory footprint.
However, for contracted Gaussians the footprint of the target VRR1 integrals may grow substantially; even though the batching over primitives can be used to reduce the memory footprint, doing so increases the cost of indexing and reduces parallelism.

\begin{figure}[ht!]
\includegraphics[width=\columnwidth]{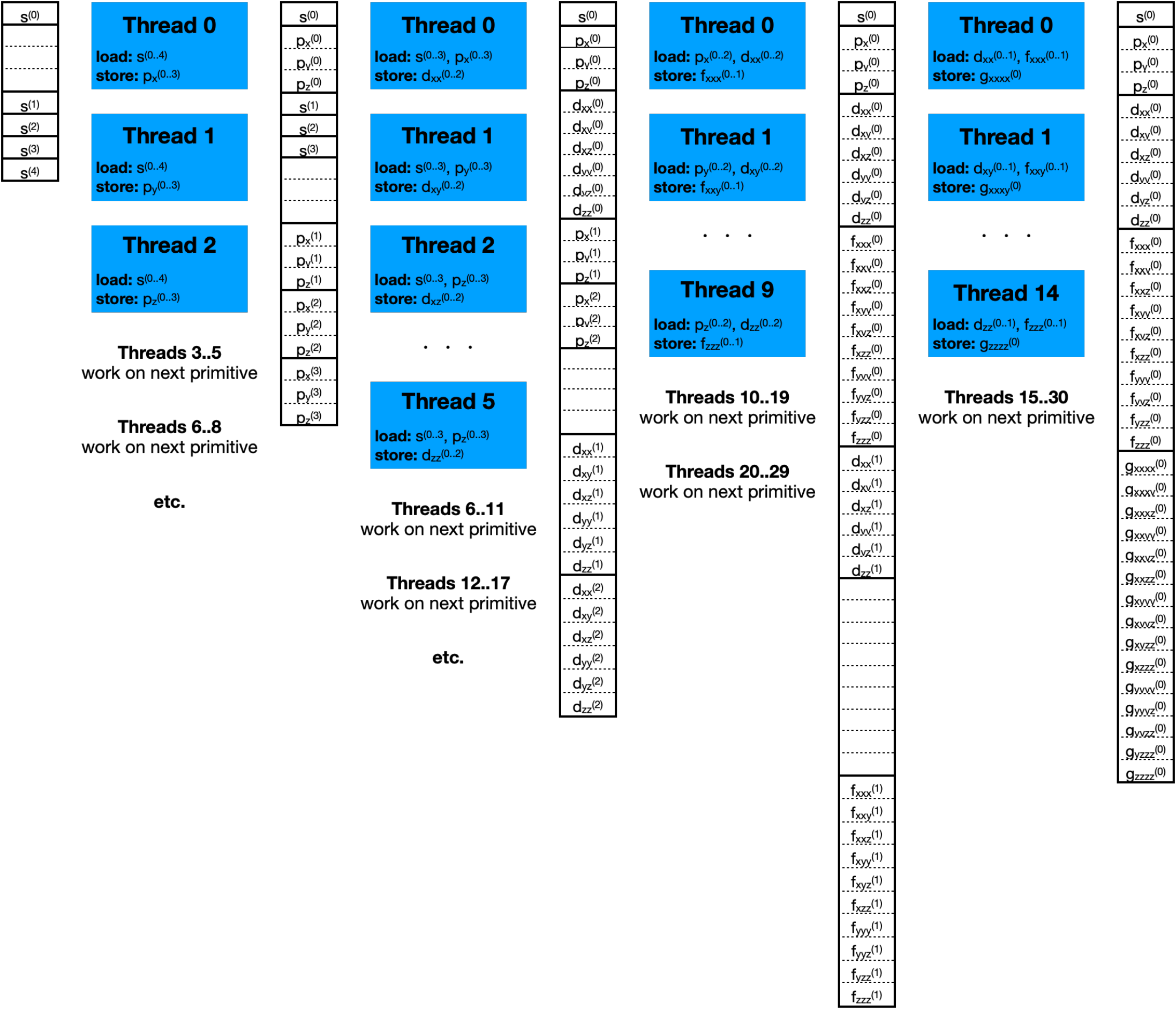}
        \caption{Mapping of the work and data flow onto the device compute resources and shared memory within the VRR1 part of the kernel for computing $(dd|s)$ using a 1-q VRR1 variant with a simple static memory allocation strategy. Specifically, shared memory layout adopted during generation of $[e]$ shell sets writes the $[e]^{(m\geq c+1)}$ shellsets to a block of memory sufficient to hold the target $[e]^{(c)}$ shell set.}
        \label{fig:vrr1-1q}
\end{figure}

A significant factor driving VRR1 memory overhead is memory management. It is easy to verify that the peak memory used by straightforward application VRR1 is usually greater than the aggregate size of the VRR1 target integral sets.
Consider evaluation of all $[e]^{(m)} \equiv [e\, 0|0]^{(m)}$ shell-sets required for VRR2, namely $[e]^{(c)}$, $|a-c| \leq e \leq a+b$. The VRR1 evaluation of $[e]^{(c)}$ uses 4 sets of integrals, namely $[e-1]^{(c)}$, $[e-2]^{(c)}$, $[e-1]^{(c+1)}$ and $[e-2]^{(c+1)}$. The latter two prerequisites are not among the VRR1 targets; ideally their storage can be reused as soon as they are not needed. Unfortunately the non-target prerequisites of $[e]^{(c)}$ occupy more space than $[e]^{(c)}$ itself whenever $e>4$. To be able to fully reuse the space occupied by the non-target prerequisites would require remapping the layout of intermediate shell-sets after each stage of applying VRR1, which would degrade performance. A simple uniform static memory management strategy for the VRR1 is not possible. \cref{fig:vrr1-1q} illustrates one particular static memory allocation strategy that we considered; however, it is not simple or uniform and leads to greatly increased memory footprint than the ideal value, as demonstrated in \cref{tab:vrr1}.

\begin{figure}[ht!]
\includegraphics[width=0.7\columnwidth]{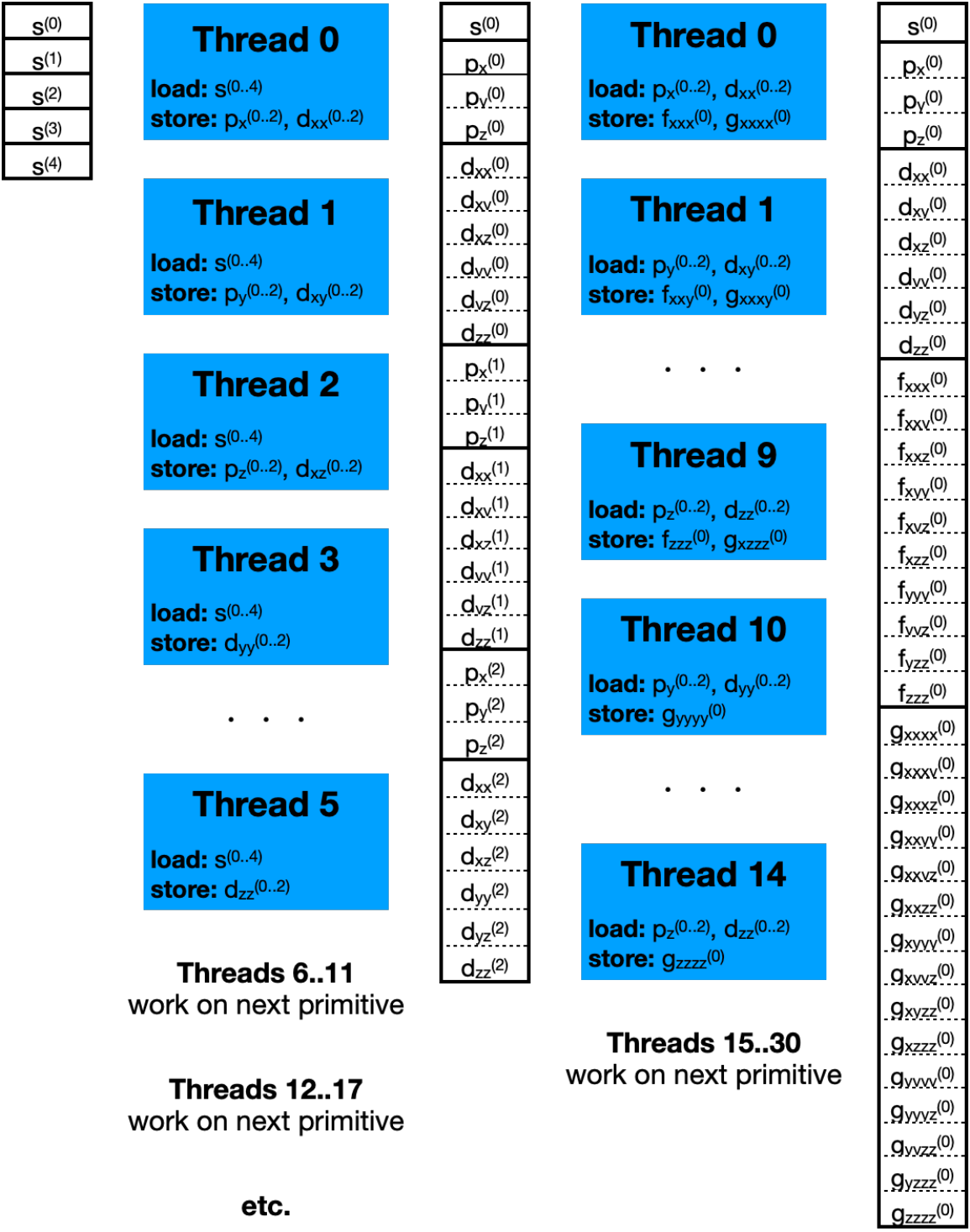}
        \caption{Mapping of the work and data flow onto the device compute resources and shared memory within the VRR1 part of the kernel for computing $(dd|s)$ using the 2-q VRR1 variant.}
        \label{fig:vrr1-2q}
\end{figure}

It turns out that it is possible to design a simple static memory management strategy and attain smaller-than-idealized footprint by evaluating target shell-sets with 2 angular momenta,
$[e]$ and $[e-1]$ (the auxiliary indices are omitted for simplicity), from the prerequisite $[e-2]$ and $[e-3]$ sets (for odd $l_\ga+l_\gb$ 2-q VRR1 applied starting with $[0]$ and $[1]$ sets, i.e., after a single round of 1-q VRR1). Although the FLOP count is slightly higher in the 2-q variant of VRR1, the memory management and indexing is greatly simplified (as illustrated in \cref{fig:vrr1-2q}) and more importantly the memory footprint is reduced even below the idealized (i.e., non-attainable) 1-q VRR1 value. Comparison to a practical 1-q memory management is even more favorable for the 2-q variant; a $~50\%$ footprint reduction is expected for building a $(ii|$ bra.

\begin{table}[ht!]
    \centering
    \setlength\tabcolsep{1.5pt}
    \begin{tabular}{c|ccc}
       \hline\hline
       $l_\ga+l_\gb$ & 1-q (ideal) & 1-q (simple)  & 2-q 
       \\ \hline
2 & 10 & 13 & 10 \\
 3 & 20 & 26 & 20 \\
 4 & 36 & 45 & 35 \\
 5 & 60 & 71 & 56 \\
 6 & 92 & 110 & 85 \\
 7 & 133 & 173 & 128 \\
 8 & 184 & 267 & 182 \\
 9 & 248 & 398 & 248 \\
 10 & 327 & 575 & 327 \\
 11 & 420 & 808 & 420 \\
 12 & 528 & 1108 & 528 \\
 13 & 652 & 1487 & 652 \\
 14 & 796 & 1958 & 793 \\
 15 & 962 & 2535 & 952 \\
 16 & 1148 & 3233 & 1131 \\
 17 & 1355 & 4068 & 1344 \\
 18 & 1584 & 5057 & 1580 \\
 \hline\hline
 \end{tabular}
 \caption{Estimated peak memory usage (in FP64 words) by 1-q and 2-q variants of VRR1 evaluation of prerequisites for the $(ab|0)$ target. ``Ideal'' refers to the 1-q variant in which all intermediates are maximally compacted after every subkernel. ``Simple'' refers to a 1-q variant with simple memory allocation  and described in \cref{fig:vrr1-1q}. The 2-q variant refers to the algorithm described in \cref{fig:vrr1-2q} (for odd $l_\ga+l_\gb$ 2-q transfers follow single 1-q transfer). Both ``1-q(simple)'' and ``2-q'' variants assume in-place evaluation.}
    \label{tab:vrr1}
\end{table}

As illustrated in \cref{fig:vrr1-2q}
depicting the mapping of the work and data onto the threads and shared memory, each VRR1 subkernel thus builds up to 2 angular momentum quanta at a time, from $[e-3]$ and $[e-2]$ to $[e-1]$ and $[e]$. Although only some subkernel threads end up writing out  $[e]$ integrals to the shared memory, internally the subkernels are implemented in terms of 1-q VRR1, i.e. , each thread evaluates $[e-1]$ even if it does not write them out.
Note that the target VRR1 integral sets end up written as a contiguous sequence at the beginning of the shared memory work region  without the need to pre-reserve the memory for the target sets; the memory management logic of 2-q is thus nearly optimal from the standpoint of simplicity and performance.  As an added benefit, the number of synchronization points is also reduced in the 2-q approach. Savings in memory, memory operations and synchronization points make up for the extra floating point operations involved in the 2-q approach.
One could extend the 2-q VRR1 idea further to transfers of even more quanta. However, since the number of recomputed intermediates
grows rapidly, at some point the FLOP increase will outweigh the benefits.  We are
investigating alternative direct methods that require fewer FLOPs.

Note how both 1-q and 2-q approaches illustrated in \cref{fig:vrr1-1q,fig:vrr1-2q} leverage the in-place evaluation to reduce the memory footprint. For example, in the 1-q approach (\cref{fig:vrr1-1q}) the second subkernel writes the $[d]^{(0)}$ set into the memory occupied by its $[s]^{(1..3)}$ input sets. Similarly, the first 2-q subkernel in the 2-q approach (\cref{fig:vrr1-2q}) writes the $[p]^{(0)}$ and $[d]^{(0)}$ sets into the memory occupied by its $[s]^{(1..4)}$ input sets. The memory estimates in \cref{tab:vrr1} would be significantly greater for both approaches if each subkernel would have to write its outputs to memory disjoint from its inputs.

\subsubsection{Implementation of VRR2\label{sec:impl-vrr2}}

For high-$K$ integrals the maximum memory footprint occurs in the VRR2 stage. To minimize the footprint of VRR2 intermediates we employed the same idea as we just used for VRR1 and will use for HRR, namely  the multiquantal recurrence. However, instead of transfering 2 quanta at a time the approach will be taken to its extreme, with all $n=l_\gc$ needed quanta  built up at once. Such $n$-quantal ($n$-q) VRR2 approach {\em entirely} avoids storage of any intermediate values outside of the registers.

The closed formula for the VRR2 target integrals $[\ga \, \gzero | \gc]$ in the $n$-q approach can be obtained straightforwardly by recursive application of \cref{eq:vrr2}; the end result is a closed formula for $[\ga \, \gzero | \gc]$ as a linear combination of the VRR1 targets $[\ga \, \gzero | \gzero]^{(c)}$. Of course, naive evaluation of this formula would significantly increase the FLOP count due to the lack of ``reuse'' of VRR2 intermediate integrals. It turns out that it is possible to reduce the cost of $n$-q VRR2 to almost that of traditional 1-q VRR2 even for the $K=1$ case by switching to the {\em contracted} version of VRR2 (CVRR2) described below; introduction of CVRR2 can be viewed in the $n$-q VRR2 context as nothing but rearranging the loops and optimizing the order of evaluation. This naturally allows to reduce the cost of VRR2 greatly for $K>1$ by moving the contraction loop inside the $n$-q VRR2 evaluation. In practice all primitive VRR1 integrals are usually available at once\footnote{This is actually a key reason to maximally reduce the VRR1 memory footprint so that batching over primitives can be avoided if at all possible.} hence the batching of the primitive sum can be mostly avoided. 
The end result is that each contracted VRR2 target $(\ga \, \gzero | \gc)$ is computed efficiently and without any extra storage.
Note that the existing uses of early contraction approaches were introduced primarily to reduce the operation cost of high-$K$ integrals\cite{VRG:pople:1978:JCP,VRG:gill:1991:IJQC}. Our use of early contraction was motivated primarily by its amenability for memory-optimal $n$-q VRR2 formulation FLOP-efficiently even for primitive ($K=1$) integrals, but the increased efficiency for contracted ($K>1$) integrals is a great bonus as well.

To arrive at CVRR2 we start by inserting the following auxiliary primitive CVRR2 integral,
\begin{align}
  \label{eq:a0cv2}
    [\ga \, \gc]_{k,\gn} \equiv & \left( \frac{ \rho}{2\gamma\zeta_c}  \right)^k \left( \left( \frac{\rho}{\zeta_c}\right)^{|\gn|} \prod_{i=x,y,z} (\mathbf{PQ})^{n_i}_i \right) [\ga \, \gzero |\gc]^{(k+|\gn|)},
\end{align}
into VRR2 (\cref{eq:vrr2}), yielding the ``primitive'' version of CVRR2:
\begin{align}
\label{eq:cVRR2_uncontracted}
    [\ga \, \gc + \bm{1}_i]_{k,\gn} = &  [\ga \, \gc]_{k,\gn+\bm{1}_i} + a_i [\ga - \bm{1}_i \, \gc]_{k+1,\gn}.
\end{align}
Since \cref{eq:cVRR2_uncontracted} does not include exponent-dependent factors, it applies to the contracted counterpart of the auxiliary CVRR2 integral \eqref{eq:a0cv2}:
\begin{align}
\label{eq:cVRR2}
    (\ga \, \gc + \bm{1}_i)_{k,\gn} = &  (\ga \, \gc)_{k,\gn+\bm{1}_i} + a_i (\ga - \bm{1}_i \, \gc)_{k+1,\gn};
\end{align}
This is the CVRR2 equation that can be applied outside of the contraction loop.
Recursive application of this relation yields its $n$-q form:
\begin{align}
    (\ga \, \gzero | \gc) \equiv (\ga \, \gc)_{0,\gzero} \overset{\text{CVRR2}}{=} & (\ga \, \gc - \bm{1}_i)_{0,\bm{1}_i} + a_i (\ga - \bm{1}_i \, \gc - \bm{1}_i)_{1,\gzero}
    \overset{\text{CVRR2}}{=} \dots \nonumber \\
    = & \sum_{\gtt \leq \gc} \sum_{\gu \leq \min(\gtt,\ga)} \left( \prod_{i=x,y,z} \begin{pmatrix}a_i \\ u_i\end{pmatrix}^2 u_i! \right) (\ga - \gu \, \gzero)_{(|\gc|+|\gu|-|\gtt|)/2,\gtt}
\end{align}
Thus each CVRR2 target is expressed as a linear combination of CVRR2 prerequisites, which are contracted versions of \cref{eq:a0cv2}.

Let us consider a concrete example: evaluation of shell-set $({\bf d} \, {\bf s} | {\bf d}) \equiv (\bm{2} \, \gzero | \bm{2}) \equiv (\bm{2} \, \bm{2})_{0,\gzero}$. Its CVRR2 evaluation involves the following prerequisites: $(\bm{2} \, \bm{0})_{0,\bm{2}}$, $(\bm{1} \, \bm{0})_{1,\bm{1}}$, and $(\bm{0} \, \bm{0})_{2,\gzero}$. Let's further consider evaluation of three individual integrals in the $({\bf d} \, {\bf s} | {\bf d})$ shell-set, namely $(d_{xx} \, s | d_{xx})$, $(d_{xz} \, s | d_{xz})$, and $(d_{xy} \, s | d_{xz})$:
\begin{align}
\label{eq:mqcvrr2_dxx_s_dxx}
    (d_{xx} \, s | d_{xx}) \overset{\text{CVRR2}}{=} & (d_{xx} \, p_{x})_{0,1_x} + 2 (p_{x} \, p_{x})_{1,0} \nonumber \\ \overset{\text{CVRR2}}{=} & (d_{xx} \, s)_{0,2_{xx}} + 4 (p_{x} \, s)_{1,1_x} + (s \, s)_{2,0}, \\
\label{eq:mqcvrr2_dxz_s_dxz}
    (d_{xz} \, s | d_{xz}) \overset{\text{CVRR2}}{=} & (d_{xz} \, p_{z})_{0,1_x} + (p_{z} \, p_{z})_{1,0} \nonumber \\ \overset{\text{CVRR2}}{=} & (d_{xz} \, s)_{0,2_{xz}} + (p_{x} \, s)_{1,1_{x}} + (p_{z} \, s)_{1,1_z} + (s \, s)_{2,0}, \\
\label{eq:mqcvrr2_dxy_s_dxz}
    (d_{xy} \, s | d_{xz}) \overset{\text{CVRR2}}{=} & (d_{xy} \, p_{z})_{0,1_x} + (p_{y} \, p_{z})_{1,0} \nonumber \\ \overset{\text{CVRR2}}{=} & (d_{xy} \, s)_{0,2_{xz}} + (p_{y} \, s)_{1,1_z}.
\end{align}
As clear from this example, each target CVRR2 integral has a variable number of contributions from the CVRR2 prerequisite integrals. Instead of precomputing the CVRR2 prerequisites into persistent memory and then computing each CVRR2 target integral from them we instead drive the computation by the CVRR2 prerequisites. Namely, each CVRR2 prerequisite integral is evaluated on-the-fly by contracting primitive VRR1 target integrals scaled by the appropriate exponent and geometry-dependent prefactors (see \cref{eq:a0cv2}). Then its nonzero contributions to the corresponding target CVRR2 integrals are evaluated and accumulated immediately, and the integral is discarded. No intermediate CVRR2 integrals are ever stored outside of the registers, and thus {\em the VRR2 stage has no effect on the overall memory footprint of a given integral kernel}.

Note that each prerequisite integral can contribute to multiple (but not all) target integrals, e.g. $(p_{x} \, s)_{1,1_x}$ and $(s \, s)_{2,0}$ both contribute to $(d_{xx} \, s | d_{xx})$ and $(d_{xz} \, s | d_{xz})$. Due to concurrency two different threads may update same target CVRR2 integral simultaneously, hence it is necessary in the CUDA context to use {\em atomic} floating-point accumulates.

Although the overall FLOP count of $n$-q CVRR2 is
slightly higher than that of traditional 1-q VRR1
for the $K=1$ case, such cases
represent only a small fraction of overall computations and not have
great effect on the overall performance. The zero net memory footprint of $n$-q CVRR2 makes it the preferred choice for implementation on GPUs even for the uncontracted case.
Once CVRR2 is done and Gaussian $\gc$ is transformed to $\gC$ the memory used in prior steps can be reused in HRR step.

\subsubsection{Implementation of HRR\label{sec:impl-hrr}}

Application of HRR is another major memory bottleneck, especially for
high-$L$, low-$K$ shell-sets. No matter the manner in which the HRR DAG is traversed, the maximum memory footprint greatly exceeds the footprint of the $(e 0|\gC)$ input shell-sets.  This is illustrated in \cref{tab:hrr} 
using HRR construction of the $(gg|\textswab{g})$ shell-set;
after single 1-q HRR transfer in a typical ``level-by-level'' DAG traversal
the size of intermediates increases from 1305 words to 2700 words. Naive application of HRR in which entire shell-sets of intermediate integrals would increase the memory footprint further, to 3456 words, after another 1-q HRR transfer. However, as is well known, only some intermediate integrals are needed; e.g., to construct all $(gg|$ shell-set integrals it is sufficient to only evaluate $d_{xx}$, $d_{yy}$, and $d_{zz}$ components of $\{(gd|,(hd|,(id|\}$ shell-sets. Similar savings are possible in subsequent intermediates. The end result is that the maximum memory footprint is determined by the size of intermediates after the first 1-q HRR transfer. Therefore it is possible to significantly reduce the memory footprint if we avoid the first round of HRR intermediates altogether by transferring 2 quanta at a time.

\begin{table}
    \begin{tabular}{cc|ccccccccc|cc}
       \hline\hline
      \multicolumn{2}{c|}{\diagbox[width=10em]{Stage}{$L_\text{tot}$}} & 4 & & 5 & & 6 & & 7 & & 8 & \multicolumn{2}{c}{Footprint} \\ \hline
      1-q & 2-q & & &&&&&&&& 1-q & 2-q\\ \hline
       0 & 0 & $(g\,s|$ & & $(h\,s|$ & & $(i\,s|$ & & $(k\,s|$ & & $(l\,s|$ & 1305 & 1305\\
        &  &  & \diagbox[width=0.5em]{}{} & $|$ & \diagbox[width=0.5em]{}{} & $|$ & \diagbox[width=0.5em]{}{} & $|$ & \diagbox[width=0.5em]{}{} & $|$\\
       1 & & & & $(g\,p|$ & & $(h\,p|$ & & $(i\,p|$ & & $(k\,p|$ & 2700 \\
        & &  & & & \diagbox[width=0.5em]{}{} & $|$ & \diagbox[width=0.5em]{}{} & $|$ & \diagbox[width=0.5em]{}{} & $|$ \\
       2 & 1 &  &  & &  & $(g\,d|$ & & $(h\,d|$ & & $(i\,d|$ & 1728$^a$ & 1728$^a$ \\
        & &  & & & & & \diagbox[width=0.5em]{}{} & $|$ & \diagbox[width=0.5em]{}{} & $|$\\
       3 & &  &  &  & & & & $(g\,f|$ & & $(h\,f|$ & 1620$^b$ \\
        & &  & & & & & & & \diagbox[width=0.5em]{}{} & $|$\\
       4 & 2 &  &  &  & & & & & & $(g\,g|$ & 2025 & 2025 \\
       \hline\hline
    \end{tabular}
    \caption{Detailed breakdown of memory footprints (in real words) of intermediates involved in a level-by-level traversal of the 1-q and 2-q HRR DAGs for evaluation of a $(gg|\textswab{g})$ shell-set.}
    \label{tab:hrr}
        $^a$ If computing only 3 nonredundant components of the $d$ shell sufficient to generate the entire Cartesian $g$ shell via 2-q HRR; memory footprint would be 3456 words, if all 6 Cartesian components of the $d$ shell were kept.\\
    $^b$ If computing only 6 nonredundant components of the $f$ shell sufficient to generate the entire Cartesian $g$ shell via 1-q HRR; memory footprint would be 3240 words, if all 10 Cartesian components of the $f$ shell were kept.
\end{table}

The 2-quantal in-place HRR algorithm is similar to our VRR1, except for the odd number of quanta the 1-q HRR stage {\em follows} the 2-q HRR stages (this is important). As \cref{tab:hrr} illustrates, the use of 2-q HRR approach avoids the storage of largest intermediates on the HRR DAG. Just as in the VRR1 case, the in-place evaluation is crucial for optimal reduction of the memory footprint.
\cref{tab:hrr-l} illustrates how the memory savings vary with the quantum numbers of the Gaussians; note that in both approaches the final (target) integrals can be written to the main memory directly and thus are not included in the reported memory footprint.
The savings are significant for both low and high angular momenta (which allows greater concurrency); e.g., for the $(ii|\textswab{i})$ shell-set the use of 2-q HRR reduces memory footprint by $\sim29\%$, which permits (in the FP64 case) for the HRR stage to fit comfortably into the 96 KiB shared memory per SM on the V100 card.
Although the use of 2-q HRR increases the FLOP count, by around 30\%, the
memory savings and fewer indexing and memory operations lead to an overall performance improvement.

\begin{table}[ht!]
    \centering
    \setlength\tabcolsep{1.5pt}
    \begin{tabular}{c|ccc}
       \hline\hline
       L & 1-q$^a$ & 2-q & Savings (\%) 
       \\ \hline
 2 & 240 & 180 & 25.0 \\
 3 & 966 & 700 & 27.5 \\
 4 & 2700 & 1728 & 36.0 \\
 5 & 6105 & 4851 & 20.5 \\
 6 & 12012 & 8502 & 29.2 \\
 \hline\hline
 \end{tabular}
 \caption{Estimated peak memory usage (in real words) by 1-q and 2-q variants of HRR evaluation of prerequisites for the $(L\, L|\textswab{L})$ target. Both ``1-q'' and ``2-q'' variants assume in-place evaluation.}
    \label{tab:hrr-l}
    $^a$ As elaborated in text, the maximum memory footprint in the 1-q approach is determined by the aggregate size of $(L \, 1|\textswab{L}) \dots (L-1 \, 1|\textswab{L})$ intermediates.
\end{table}

\section{General Aspects of the Implementation}\label{sec:implementation}

It is appropriate to  elaborate on how modern C++17 compile-time programming techniques are actually deployed. Due to the space limitations, such discussion will be kept brief. We refer the reader to the openly-available library source for more details.

For example, consider the problem of parallel recurrence algorithms on GPU. To evaluate the target integral assigned to it, a given
thread needs to compute the address of the corresponding input integrals as well as the recurrence prefactors from the knowledge of its thread index in the thread block. Such index computation is complex, hence it makes sense to precompute index maps in constant tables rather than computing in each thread at the runtime.
The \code{constexpr} functions and compile-time control flow, such as \code{if constexpr} statements, allow for generating such
tables in C++17 with traditional syntax, rather than cumbersome template metaprogramming. The compile-time generation of such metadata also allows their  placement in constant device memory directly, without the need to manage host-device transfers at runtime.


More importantly, the use of \code{constexpr} metadata is critical for being able to express irregular recurrences generically, as a single function by the quantum numbers of the Gaussians,
since the compiler can then to unroll the (fixed-size) loops at the template instantiation time using \code{constexpr} metadata such as lists of the quantum numbers of Cartesian AOs in shells (compile-time generation of such lists in C++ is illustrated in \cref{lst:constexpr-orbital-list}), elide zero contributions to the target integrals, perform aggressive common subexpression elimination, and other optimizations that normally a custom code generator would have to do.



The end-user API of the library is plain C++; no device API artifacts (e.g., streams) are visible to the user.  The AO integral batches do not need to be in any particular order.
The low-level device-level API, while not intended for end users, is also available for expert use and customization. For example, the CUDA integral evaluation kernels accept as a template parameter a {\em callable} (e.g., a lambda function) that performs integral digestion; this allows to post-process integrals without writing them into the
global memory of the device. This mechanism, for example, was used to implement a simple
density-fitted J-matrix construction.  The DF-J builder is exposed to the end user through the device-independent C++ interface that receives tiles of the input (density) matrix and writes the tiles of the output (J) matrix via \code{std::function} to avoid making any assumptions regarding data layout or multi-process execution.

\begin{center}
\begin{listing}[H]
\begin{minted}{c++}
#include <array>
#include <utility>

#define LMAX 3

struct Orbital {
  int x,y,z;
};

constexpr auto orbital(int idx) {
  return Orbital{
    // x,y,z orbital corresponding to index idx
  };
}

template<int ... Is>
constexpr auto make_orbitals(std::integer_sequence<int,Is...>) {
  return std::array<Orbital,sizeof...(Is)>{
    orbital(Is)...
  };
}

constexpr auto orbitals = make_orbitals(
  std::make_integer_sequence<int, ((LMAX+1)*(LMAX+2)*(LMAX+3))/6>()
);
\end{minted}
\caption{Compile-time generation of lists of Cartesian AO quanta in shells.}
\label{lst:constexpr-orbital-list}
\end{listing}
\end{center}

\section{Performance}\label{sec:performance}

To assess the performance of the integral kernels in \code{LibintX} we
compared the \code{LibintX} CPU and GPU kernels against \code{Libint}'s (CPU-only) kernels for a representative subset of shell-sets and three values of total contraction degree ($K=1,5,25$). We also compared
the performance of \code{LibintX} for the evaluation of the Coulomb potential (J matrix) against its \code{Libint}-based CPU-only implementation in MPQC\cite{VRG:peng:2020:JCP}.
Computations were carried out on a node with 2 Intel Xeon Gold 6136 CPUs (24 cores total)
and NVIDIA V100 GPUs. The CPU code was compiled with gcc 10.3.0 and \code{-mtune=native -march=native -Ofast} optimization flags. To simplify benchmarking the CPU-based kernels were launched on a single CPU core when profiling the performance for individual shell-sets. Such comparison should be viewed as a very conservative estimate of the relative GPU-vs-CPU performance, by neglecting the effects of L2 cache contention by CPU cores as well as minimizing thermal effects on the CPU performance. Conservative estimates of performance speedup of 1 GPU vs 1 12-core CPU can be obtained by dividing the values in \cref{tab:abc-gpu} by 12.
In profiling a more realistic application, namely the evaluation of Coulomb potential, all 24 cores of the 2 CPUs in a node were used.

The overall CPU performance of \code{LibintX} is a moderate to substantial improvement (with a few exceptions) over \code{Libint} (see \cref{tab:abc-cpu})
validating that the C++ compiler is able to optimize generic kernels in \code{LibintX} well. Note, however, that only the Boys function evaluation in \code{Libint} is vectorized (using AVX intrinsics), thus the comparison may appear actually rather favorable to \code{Libint}, especially for the high $L$ shell-sets, where most of the time is spent in scalar (non-SIMD) portions of the \code{Libint} kernels. However, the better performance of \code{LibintX} is actually promising considering the relative simplicity of its implementation. A few more noteworthy conclusions can be drawn.
\begin{itemize}
\item The substantial performance improvement for the $(ss|s)$ class suggests better quality of implementation of the Boys engine and/or primitive data evaluation in \code{LibintX}.
\item For classes dominated by VRR1, e.g., $(l0|0)$, the \code{LibintX} performance advantage seems to decrease with $l$. It appears that the FLOPS increase due to the use of multiquantal VRR1 is difficult to overcome by better vectorization.
\item For classes dominated by HRR, e.g., $(ll|0)$, the advantage of \code{LibintX} approaches $\sim2$ for high $l$, but for some reason drops for $l=2,3$.
\item For classes dominated by VRR2, e.g., $(l0|l)$, the advantage of \code{LibintX} approaches is consistent and increases steeply with $K$, likely due to the use of early-contraction variant of VRR2 in \code{LibintX}. For $(ll|l)$ classes with high $K$ and high $L$ the advantage of \code{LibintX} reaches as high as $17$.
\end{itemize}

\begin{table}
    \centering
    \begin{tabular}{c|ccc}
    \hline\hline
    $l_\ga l_\gb l_\gc$ & $K=1$ & $K=5$ & $K=25$  \\ \hline
    000 & 2.00 & 1.90 & 1.73 \\
100 & 5.29 & 2.41 & 1.94 \\
200 & 2.76 & 1.89 & 1.51 \\
300 & 2.10 & 1.31 & 1.06 \\
400 & 1.63 & 1.30 & 1.14 \\
500 & 1.43 & 1.22 & 1.11 \\
600 & 1.33 & 1.19 & 1.11 \\
101 & 2.12 & 1.69 & 1.39 \\
202 & 1.45 & 1.23 & 1.09 \\
303 & 1.07 & 1.41 & 1.46 \\
404 & 1.68 & 3.01 & 3.87 \\
505 & 1.48 & 4.12 & 6.42 \\
606 & 1.57 & 5.21 & 10.04 \\
110 & 2.12 & 1.80 & 1.50 \\
220 & 1.39 & 1.03 & 0.88 \\
330 & 1.04 & 0.82 & 0.75 \\
440 & 1.79 & 1.59 & 1.41 \\
550 & 2.07 & 2.06 & 2.04 \\
660 & 2.08 & 2.06 & 2.04 \\
111 & 1.71 & 1.46 & 1.26 \\
222 & 1.46 & 2.00 & 2.48 \\
333 & 1.75 & 3.10 & 5.15 \\
444 & 2.17 & 5.24 & 10.74 \\
555 & 2.26 & 6.21 & 14.09 \\
666 & 2.67 & 7.03 & 17.24 \\
    \hline\hline
    \end{tabular}
    \caption{Relative performance of CPU-based \code{LibintX} kernels vs. the reference \code{Libint} counterpart on 1 CPU core.}
    \label{tab:abc-cpu}
\end{table}


Comparison of \code{LibintX}'s performance on 1 V100 GPU with its performance on a single CPU core is presented in \cref{tab:abc-gpu}.
The baseline for the comparison is the $\sim 80$ ratio of peak FLOPS throughout of the entire V100  GPU ($\sim 7.8$ FP64 TFLOPS) to that of a single CPU core ($\sim 96$ FP64 GFLOPS using AVX-512 multiply-add instructions. In addition we reported the FLOP rates (as a percentage of hardware peak) as reported by the NVIDIA Nsight profiler. Several conclusions can be drawn from the data.
\begin{itemize}
\item The low speedups and low FLOP rates for low-$L$ classes with $K=1$ [e.g., $(ss|s)$ and $(ps|s)$] are partially due to the insufficient work per thread block. Namely, recall that we assign single contracted shell-set to a thread-block. Thus for the $(ps|s)$ shell-set with $K=1$ only 3 lanes of a warp will be occupied, with the rest of warp lanes left unused. To improve performance for such classes we would need to assign multiple shell-sets to each thread block, which would complicate the implementation substantially. Since we expect that the computational cost of such integrals will be rather low with realistic bases we did not at this time pursue this optimization, but will consider it in the future. 
\item To reach adequate performance on the GPU each thread block needs to contain enough work for at least 2 warps (64 threads). For example, the $(gs|s)$ shell-set (consisting of $15\times1\times1$ integrals) with $K=5$ uses $15\times5=75$ threads.
\item In general the performance increases with $K$, except for very high $L$, where the
pressure on the shared memory decreases the performance.
\item In practical applications we expect the costs to be determined by primitive integrals moderate to high angular quanta, such as $(ff|f)$, and/or moderate angular momenta with nonunit contraction degree, such as $(ds|d)$ or $(pp|p)$. For those cases the
performance almost always matches of beats
the target threshold. 
\item The measured FLOP rates at first glance seem disappointing: only a few percent with $K=1$ (as high as $7\%$ for the $(ff|f)$ class) and moderately better for $K=5$ and $K=25$, reaching as high as $14\%$ for the $K=25$ $(dd|s)$ shell set. Unfortunately, there are not too many integral kernel microbenchmarks reported in the literature; most performance results in the literature report total execution times for complete applications.
For 4-center Coulomb integrals we can compare these results to the microbenchmarks reported by Barca et al. in Ref. \citenum{VRG:barca:2021:JCTC} for 4-center integrals implemented using the OS-based HGP scheme. For the $(ds|ss)$ class (using a mixture of shell sets supported by the 6-31G(d) basis, i.e. with $K$ between 1 and 36) they reported performance of ~3 TFLOPS, or $~38\%$ of the peak. The counterpart in our work is the $(ds|s)$ class for which we obtained FLOP rates of between $1\%$ and $12\%$, which is substantially slower than that reported in Ref. \citenum{VRG:barca:2021:JCTC}. On the other hand, the performance reported by Barca for larger shell-sets is significantly lower: for the uncontracted $(dd|dd)$ class they reported FLOP rate of 151 GFLOPS (or $\sim2\%$ of the peak), largely limited by the excessive memory footprint of the recursion. A fair comparison for the $(dd|dd)$ shell-set consisting of 1296 Cartesian integrals are our uncontracted $(ff|f)$ and $(gg|g)$ shell-sets containing 1000 and 3375 solid-harmonic integrals and corresponding to FLOP rates of $7\%$ and $6\%$, respectively. Thus while there is still much room to improve the performance, for the high-$L$ classes the performance of \code{LibintX} is competitive to or exceeds that of the state-of-the-art methods. Alternative approaches to improving the performance will be reported elsewhere.
\end{itemize}

\begin{table}
    \centering
    \begin{tabular}{c|cc|cc|cc}
    \hline\hline
    $l_\ga l_\gb l_\gc$ & \multicolumn{2}{c|}{$K=1$} & \multicolumn{2}{c|}{$K=5$} & \multicolumn{2}{c}{$K=25$}  \\ \hline
    000 &   8.35 &  1\% &  28.33 &  2\% & 115.53 &    11\%   \\
100 &  10.16 &  1\% &  33.31 &  3\% & 118.51 &    11\%   \\
200 &  14.78 &  1\% &  43.48 &  4\% & 126.38 &    12\%   \\
300 &  19.81 &  1\% &  61.69 &  6\% & 139.29 &    14\%   \\
400 &  28.41 &  2\% &  77.86 &  8\% & 125.15 &    14\%   \\
500 &  37.05 &  2\% &  76.08 &  8\% &  91.91 &    11\%   \\
600 &  44.52 &  3\% &  89.11 & 10\% &  92.01 &    12\%   \\
101 &  16.74 &  1\% &  47.25 &  3\% & 159.25 &    10\%   \\
202 &  36.35 &  1\% &  92.58 &  4\% & 188.42 &    11\%   \\
303 &  62.68 &  3\% & 118.16 &  6\% & 173.43 &    10\%   \\
404 &  84.03 &  4\% & 135.22 &  7\% & 134.70 &     8\%   \\
505 &  93.63 &  4\% & 117.99 &  6\% &  79.06 &     5\%   \\
606 &  91.13 &  3\% &  81.64 &  4\% &  56.18 &     4\%   \\
110 &  16.91 &  1\% &  46.11 &  4\% & 126.45 &    12\%   \\
220 &  31.41 &  3\% &  77.30 &  8\% & 125.56 &    14\%   \\
330 &  43.24 &  4\% &  81.54 & 10\% &  73.43 &    10\%   \\
440 &  36.16 &  4\% &  61.78 &  9\% &  57.12 &     9\%   \\
550 &  30.49 &  4\% &  47.69 &  9\% &  34.34 &     7\%   \\
660 &  25.74 &  3\% &  39.35 &  7\% &  42.56 &     9\%   \\
111 &  26.22 &  1\% &  64.47 &  5\% & 158.31 &    12\%   \\
222 &  91.31 &  6\% & 132.28 & 11\% & 130.05 &    12\%   \\
333 & 103.77 &  7\% &  98.38 &  9\% &  85.49 &    10\%   \\
444 &  89.63 &  6\% &  78.22 &  7\% &  47.05 &     6\%   \\
555 &  45.36 &  3\% &  46.95 &  4\% &  41.81 &     5\%   \\
666 &  74.36 &  4\% &  21.71 &  3\% &   8.22 &     3\%   \\
    \hline\hline
    \end{tabular}
    \caption{Relative performance of CUDA \code{LibintX} kernels using one V100 card vs. the CPU-only \code{LibintX} counterpart on one CPU core along with the fraction of hardware peak FLOP achieved on the GPU. }
    \label{tab:abc-gpu}
\end{table}


To assess the performance of \code{LibintX} kernels for realistic workloads we implemented density-fitting-based Coulomb potential (J-matrix) evaluation in MPQC and compared its performance on one V100 GPU (7.8 FP64 TFLOPS peak) to that of the nearly identical reference implementation using \code {Libint} on 2 Xeon Gold 6136 CPUs (1.95 FP64 TFLOPS peak), with identical screening. The results are listed in \cref{tab:dfj}. The observed speedups are higher than one would
expect from the improved integrals performance alone; there are other factors at play here that improve performance such as better memory access patterns and tighter integration between the integral evaluation and transform steps. This is provided as an example of performance improvements that could achieve
by re-writing overall implementation {\it together} with improved GPU
integrals.

It should be noted that the J-matrix evaluation using complete integrals is known to be inefficient compared to the J-engine approach\cite{VRG:white:1994:JCP} in which the density matrix is contracted with the integral intermediates rather than with the integrals themselves. Recently Kussmann et al. reported an efficient implementation of DF-based J-engine approach for CPUs and GPUs,\cite{VRG:kussmann:2021:JCTC}
whose performance far exceeds that reported here. The techniques we developed in this paper can be productively used to the J-engine approach; the results will be reported elsewhere soon.

\begin{table}
    \centering
    \begin{tabular}{cc|cc|ccc}
    \hline\hline
    Molecule & Basis & \#AO & \#DFAO & wall time(s) & Speedup  \\ \hline
    taxol & def2-SVP        & 1,099 & 3,528  & 1.2  & 8.3 \\
taxol & def2-TZVP       & 2,185 & 3,528  & 3.4  & 6.6 \\
valinomycin & def2-SVP  & 1,542 & 4,812  & 2.4  & 12.9  \\ 
valinomycin & def2-TZVP & 2,958 & 4,812  & 6.7  & 9.4 \\ 
at4 & def2-SVP          & 2,746 & 8,876 & 7.1  & 10.6 \\ 
at4 & def2-TZVP         & 5,546 & 8,876 & 22.5 & 8.8 \\ 
olestra & def2-SVP      & 3,840 & 11,633 & 9.4  & 12.6 \\ 
olestra & def2-TZVP     & 7,093 & 11,633 & 25.1 & 8.9 \\ 

    \hline\hline
    \end{tabular}
    \caption{Performance of the DF-based Coulomb potential evaluation using \code{LibintX} kernels executing on one V100 card, and corresponding speedups vs. the \code{Libint}-based counterpart executing on 24 CPU cores. def2-universal-Jfit basis was used for density fitting.}
    \label{tab:dfj}
\end{table}

\section{Summary}\label{sec:summary}

Utilizing memory hierarchy efficiently has been always crucial for performance of high-performance algorithms, however recent development of accelerated hardware is making it even more important.
Here we reported how to re-design Gaussian integral recurrences for the accelerated architectures like modern GPGPUs by (a) using multi-quantal recurrences in favor of more traditional (and FLOP-optimal) uniquantal recurrences to reduce the memory footprint of such FLOP-efficient but memory-hungry approaches, (b) leveraging scratchpad memory (e.g., shared memory in CUDA) and the associated programming model features for further reduction of the memory footprint, and (c) implementing such recurrences without traditionally-used custom compiler/code generator by leveraging compile-time features of modern C++ compilers.
The result is an open-source library \code{LibintX}, available at \url{https://github.com/ValeevGroup/libintx}, whose implementation is described and assessed here. The benchmark performance of integral kernels on conventional and accelerated architectures match or improve on the state-of-the-art CPU implementation. For a realistic problem (Coulomb potential evaluation in representative medium-sized molecules) \code{LibintX}-based code on 1 GPU (7.8 FP64 TFLOPS peak) outperforms the conventional CPU counterpart on 24 CPU cores (1.95 FP64 TFLOPS peak) on average by an order of magnitude.
Although, like every other implementation, our particular algorithmic and implementation choices are a heuristic compromise between simplicity and performance, further refinements of these choices are sure to improve on these performance results.  Although it is likely that the optimal algorithm and implementation methods are class and platform specific, it is likely that our heuristics are competitive solutions for at least some integral classes, and their elements can be used to improve other existing heuristics.

The techniques put forth in this work are more generally applicable than just evaluation of the 3-center 2-particle Gaussian integrals. Their uses for efficient Coulomb potential evaluation in Gaussian basis as well as other contexts will be reported elsewhere soon.

\begin{acknowledgement}
This research was supported by the Exascale Computing Project (17-SC-20-SC), a collaborative effort of the U.S. Department of Energy Office of Science and the National Nuclear Security Administration.

We also acknowledge Advanced Research Computing at Virginia Tech (www.arc.vt.edu) for providing computational resources and technical support that have contributed to the results reported within this paper.
\end{acknowledgement}

\begin{appendices}

\section{Notation}
\label{sec:notation}

In this paper we follow the established notation for Gaussian integrals. 3-center 2-electron integral over (non-normalized) primitive Cartesian Gaussians are denoted by square brackets:
\begin{align}
    [\ga \gb | \gc] \equiv \iint \phi_\ga({\bf r}_1) \phi_\gb({\bf r}_1) O(|{\bf r}_1-{\bf r}_2|) \phi_\gc({\bf r}_2) \, \mathrm{d} {\bf r}_1 \, \mathrm{d} {\bf r}_2
\end{align}
with non-negative integer ``quanta'' ($\ga = \{a_x, a_y, a_z \}$, etc.) indicating also the origin and exponents of the Gaussian:
\begin{align}
    \phi_\ga ({\bf r}) \equiv (x - A_x)^{a_x} (y - A_y)^{a_y} (z - A_z)^{a_z} \exp(-\zeta_a |{\bf r} - {\bf A}|^2).
\end{align}
$l_\ga \geq 0$ will denote the sum of quanta of a Cartesian Gaussian, also (imprecisely) referred to the ``angular momentum'' of a Gaussian.

A complete set of Cartesian Gaussians that only differ in the distribution of a fixed number of quanta $l$ among the axes form a {\em shell}.
Shells with $l=0,1,2,3,4,5,6,7$ will be referred to as $s$, $p$, $d$, $f$, $g$, $h$, $i$, and $k$, respectively. A shell of Cartesian Gaussians with $l$ quanta consists of $(l+1)(l+2)/2$ elements.

$[a b|c]$ will denote a {\em shell-set} (or, simply, {\em shell}) of Gaussian integrals consisting of all integrals composed of Cartesian Gaussian shells with $l=a,b,$ and $c$, respectively.

Integrals over (non-normalized) {\em contracted} Cartesian Gaussians are denoted by parentheses, e.g. $(\ga \gb | \gc)$, with $L=l_\ga + l_\gb + l_\gc$ denoting total angular quanta of the integral and $K$ denoting the product of contraction degrees of each Gaussian; the corresponding shell-set will be denoted $(a b| c)$ with the quanta defining a particular integral {\em class}.  Solid harmonic counterpart of Cartesian Gaussian $\gc$ will be denoted by $\gC$, with $l_\gC$ denoting the angular momentum of $\gC$; the corresponding shell-set will be denoted $(a b| \gC)$.

In this work we only consider Coulomb integrals, i.e., $O$ is the Poisson kernel: $O(|{\bf r}_1-{\bf r}_2|) = |{\bf r}_1-{\bf r}_2|^{-1}$.

The following standard definitions are used throughout the manuscript:
\begin{align}
    \gamma \equiv \, & \zeta_a + \zeta_b, \\
    \rho \equiv \, & \frac{\gamma \zeta_c}{\gamma + \zeta_c}, \\
    {\bf P} \equiv \, & \frac{\zeta_a {\bf A} + \zeta_b {\bf B}}{\gamma}, \\
    {\bf AB} \equiv \, & {\bf A} - {\bf B}, \\
    {\bf PC} \equiv \, & {\bf P} - {\bf C}, \\
    F(x) \equiv \, & \int_0^1 \, \mathrm{d}y \, y^{2m} \exp(-xy^2),
\end{align}
the latter of which is the well-known Boys function.\cite{VRG:boys:1950:PRSMPES}

\end{appendices}

\bibliography{vrgrefs}

\end{document}